\documentclass[runningheads]{llncs}

\usepackage[T1]{fontenc}
\usepackage{amsmath}

\usepackage{graphicx}
\usepackage{multirow}
\usepackage[table,xcdraw]{xcolor}

\usepackage{lscape}
\usepackage{lineno}
\usepackage[utf8]{inputenc} % allow utf-8 input
\usepackage[T1]{fontenc}    % use 8-bit T1 fonts
\usepackage{hyperref}       % hyperlinks
\usepackage{url}            % simple URL typesetting
\usepackage{booktabs}       % professional-quality tables
\usepackage{amsfonts}       % blackboard math symbols
\usepackage{nicefrac}       % compact symbols for 1/2, etc.
\usepackage{microtype}      % microtypography
\usepackage[utf8]{inputenc}
\usepackage{amsmath}
\usepackage{amssymb}

\usepackage{stfloats}

\usepackage{bm}
\usepackage[algoruled, longend, boxed]{algorithm2e}
\usepackage{mathrsfs}
\usepackage{float}
\usepackage{subfigure}
\usepackage{siunitx}
\usepackage{xcolor}
\DeclareUnicodeCharacter{00A0}{~}
\nolinenumbers

\SetKwInput{KwInput}{Input}                % Set the Input
\SetKwInput{KwOutput}{Output}              % set the Output
\SetKwInput{KWInitialization}{Initialization}                % Set the Input
\SetKwInput{KWReturn}{Return}                % Set the Input

\begin{document}
\title{Predicting conversion of mild cognitive impairment to Alzheimer's disease by modelling healthy ageing trajectories\thanks{Under review}}
\titlerunning{Predicting conversion of mild cognitive impairment to Alzheimer’s disease}
% If the paper title is too long for the running head, you can set
% an abbreviated paper title here
%
\author{Yiran Wei\inst{1} \and  Stephen J. Price\inst{1} \and Carola-Bibiane Schönlieb\inst{2}  \and  Chao Li\thanks{Corresponding author}\inst{1,2}} 
% \and
% Second Author\inst{2,3}\orcidID{1111-2222-3333-4444} \and
% Third Author\inst{3}\orcidID{2222--3333-4444-5555}}
%
\authorrunning{Y. Wei et al.}
% First names are abbreviated in the running head.
% If there are more than two authors, 'et al.' is used.
%
\institute{Department of Clinical Neurosciences, University of Cambridge  
\and Department of Applied Mathematics and Theoretical Physics, University of Cambridge}
\maketitle              % typeset the header of the contribution
\begin{abstract}
Alzheimer's disease (AD) is the most common age-related dementia. Mild cognitive impairment (MCI) is the early stage of cognitive decline before AD. It is crucial to predict the MCI-to-AD conversion for precise management, which remains challenging due to the diversity of patients. Previous evidence shows that the brain network generated from diffusion MRI promises to classify dementia using deep learning. However, the limited availability of diffusion MRI challenges the model training. In this study, we develop a self-supervised contrastive learning approach to generate structural brain networks from routine anatomical MRI under the guidance of diffusion MRI. The generated brain networks are applied to train a learning framework for predicting the MCI-to-AD conversion. Instead of directly modelling the AD brain networks, we train a graph encoder and a variational autoencoder to model the healthy ageing trajectories from brain networks of healthy controls. To predict the MCI-to-AD conversion, we further design a recurrent neural networks based approach to model the longitudinal deviation of patients' brain networks from the healthy ageing trajectory. Numerical results show that the proposed methods outperform the benchmarks in the prediction task. We also visualize the model interpretation to explain the prediction and identify abnormal changes of white matter tracts. 

\keywords{Brain networks \and Graph neural networks \and Alzheimer's disease.}
\end{abstract}
\section{Introduction}
\subsection{Alzheimer's disease and mild cognitive impairment}
Alzheimer's disease (AD) is the most common cause of dementia, characterized by continuous decline in cognition, memory and brain functions \cite{mattson2004pathways}. Mild cognitive impairment (MCI) is regarded as the intermediate stage of cognitive decline between healthy ageing and AD, where a proportion of MCI cases could be reversible. Further, although lacking curative treatment, earlier identification and intervention could modify the disease trajectory of AD progression and impact patient outcomes. Therefore, it is of crucial significance to accurately predict the conversion from MCI to AD, which remains a significant challenge due to the heterogeneous nature of AD\cite{au2015back}. 

\subsection{Neuroimaging and brain networks}
Magnetic resonance imaging (MRI) is a commonly-used noninvasive technique for managing neuropsychiatric conditions. Previous studies suggest that MRI can detect the structural change of the brain in dementia patients \cite{elliott2020mri}. The MRI-derived biomarkers are reported significantly associated with cognitive decline, indicating the clinical value of MRI in dementia prediction\cite{cox2019structural}

The structural brain network, constructed from diffusion MRI (dMRI) or anatomical MRI, is promising to characterize the connectivity between cortical/subcortical regions defined according to prior knowledge of neuroanatomy. Specifically, dMRI-derived brain networks utilize tractography to quantify the connectivity strength of tracts that link brain regions. At the same time, anatomical MRI examines the possible associations among brain regions by calculating the covariance of the anatomical features (e.g., grey matter volume) among brain regions. Both types of structural brain networks are reported to effectively generate useful graph theoretical biomarkers associated with various neuropsychiatric conditions \cite{griffa2013structural,wei2021quantifying,wei2021structural} including dementia \cite{ajilore2014association}. Although dMRI provides more direct connectivity estimation than anatomical MRI, it is more difficult to acquire. In parallel to the efforts in developing data augmentation approaches \cite{li2021brainnetgan,barile2021data}, there is a pressing need to generate robust brain networks from more commonly available anatomical MRI.

\subsection{Deep learning for neuroimaging}
Deep learning models demonstrate reasonable performance in classifying AD and predicting MCI-to-AD conversion based on neuroimaging ~\cite{abrol2020deep}. However, the majority models are developed for computer vision tasks, which may not capture the connectomic properties of the brain. Therefore, the model performance could be limited for complex tasks involving domain knowledge, e.g., predicting disease trajectory of AD. Moreover, the interpretability of these models could be further improved to facilitate clinical translation. In parallel, the emerging graph neural network (GNN), designed for learning non-Euclidean data, is widely used in characterizing neuropsychiatric diseases \cite{schirmer2021neuropsychiatric}, which has shown encouraging performance in AD research, e.g., tau spread networks, brain structure geometric, based on MRI-derived graph structure data  ~\cite{sarasua2021transformesh,song2020physics}.

\subsection{Related work}
In general, the approaches of predicting the  MCI-to-AD conversion are categorized as static models (based on single time point data) or dynamic models (based on longitudinal data). Static models predict whether the conversion will happen in 36 months solely based on the baseline images~\cite{abrol2020deep,gao2020ad}, which, however, ignores the longitudinal changes along the disease trajectory. Therefore, although such approaches require fewer data, they could be sub-optimal due to the static nature of the predicting models.
In contrast, dynamic models are emerging approaches to predict MCI-to-AD conversion, particularly with the availability of longitudinal datasets, e.g., the Alzheimer's Disease Neuroimaging Initiative (ADNI). A recent study applies GNN and recurrent neural networks (RNN) to predict the patient outcomes in 18-month follow up by modeling the longitudinal brain networks from the baseline to 12-month as dynamic graphs  \cite{kim2021interpretable}. However, this method showed limited performance, which might be due to its attempts to directly model the heterogeneous dementia population using the end-to-end training scheme. Moreover, the input brain networks are constructed solely from grey matter features, which ignores the common white matter abnormalities in AD. 

\subsection{Proposed framework}
Previous studies show that anatomical MRI and dMRI share common features in reflecting brain structure \cite{alexander2013convergence,gu2019generating}. Hence, it could be feasible to generate brain networks from anatomical MRI with the guidance of dMRI that provides more specific information regarding white matter tracts. In this way, we could fully characterize both grey matter and white matter of the brain. 
In addition, accumulating research shows that AD patients demonstrate accelerated brain ageing compared to healthy controls (CN) and MCI patients \cite{cole2018brain,gao2020ad}. Therefore, we hypothesize that the MCI-to-AD conversion could be predicted by modelling the deviation of AD patients from the healthy ageing trajectory, which could mitigate the challenge of modelling brain networks in the heterogeneous AD population. 

Here we propose a learning framework to model the healthy ageing trajectory of brain networks with a graph encoder and a variational autoencoder (VAE). In addition, we design an RNN based algorithm that predicts the MCI-to-AD conversion based on the past longitudinal deviations/residuals of patients from the predicted ageing trajectory. In order to generate brain networks from commonly available anatomical MRI, we propose a self-supervised approach that uses an autoencoder to extract node features from T1 images and a cross-modal contrastive representative learning approach to extract edge features guided by dMRI. Our contributions include:
\begin{itemize}
    \item A cross-modal learning approach to generate brain networks by extracting features from anatomical MRI under the guidance of dMRI. 
    \item A generative approach to predict the healthy ageing trajectories of brain network features using graph neural networks and variational autoencoders.
    \item A recurrent learning algorithm that models longitudinal residuals between patient's actual features and patient's predicted features from the healthy ageing trajectory for future disease status prediction.
    \item An interpretation approach that identifies the abnormality introduced from MCI-to-AD conversion by comparing the actual diseased brain networks and the predicted healthily aged brain networks from a customized graph decoder.

\end{itemize}

\section{Methods}
\subsection{Data preparation}
A longitudinal MRI dataset of AD, MCI and CN subjects is downloaded from ADNI website. Each subject has one T1 images at baseline, 6 months, 12 months and 18 months respectively. In total, the longitudinal dataset includes 191 stable CN, 126 stable MCI, and 91 converted MCI. 
Another independent baseline cohort including 113 CN and 96 AD with dMRI and T1 available are also downloaded from the ADNI website. 

T1 images of all subjects are transformed to the standard MNI-152 space by coregistrating with MNI-152 standard T1 of the FMRIB Software Library (FSL) using Advanced normalization tools \cite{avants2009advanced,jenkinson2012fsl}. For the dMRI of the independent baseline cohort, fractional anisotropy (FA) maps are derived using the FMRIB's Diffusion Toolbox and transformed to the standard space by coregistrating with the standard FA map.

\subsection{Brain network construction}
\begin{figure}[hbt!]
\centering
\includegraphics[width=0.8\textwidth]{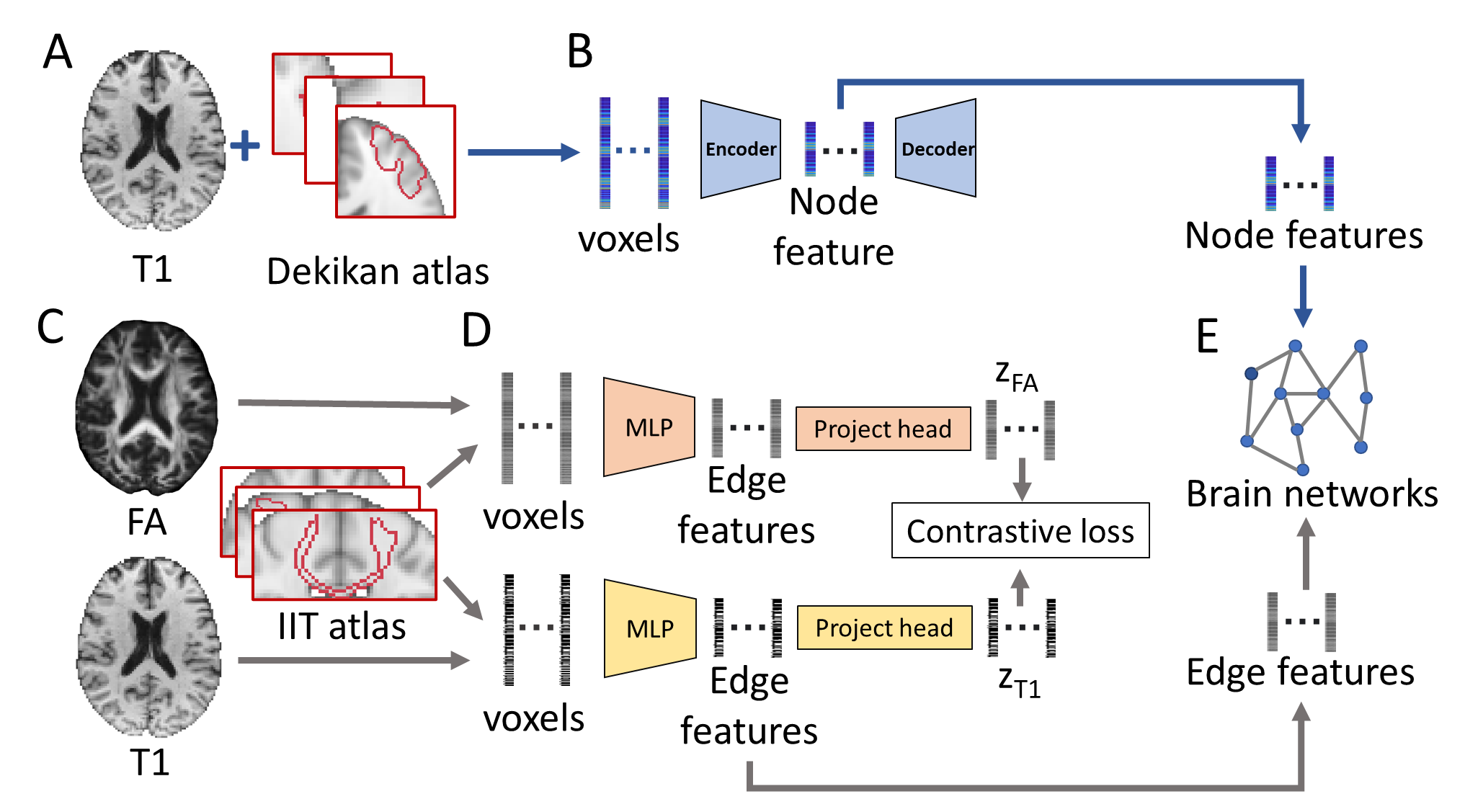}
\caption{Workflow of generating brain networks. \textbf{A.} T1 image and Dekikan grey matter node atlas are combined to obtain voxel vectors of node regions. \textbf{B.} Node voxel vectors are fed into an autoencoder to reduce dimensionality and produce node features for brain networks. \textbf{C.} Voxels of FA and T1 enclosed by the IIT white matter atlas are extracted as the edge voxel vectors. \textbf{D.} A cross-modal contrastive learning model is used to extract FA-related T1 features corresponding to edges.  \textbf{E.} Node features and edge features are arranged into graph format for the downstream training.} 
\label{fig:brainnetworks}
\end{figure}

Node and edge features of brain networks are both learnt from the independent baseline data. (Fig~\ref{fig:brainnetworks}). The Desikan grey matter atlas is used as the node atlas that divides cortical and sub-cortical regions into 68 separate areas with cerebellum excluded (Fig~\ref{fig:brainnetworks}A). For each subject, the Desikan node atlas is transformed back to the native space of the subjects using the inverse coregistration file from data preparation. The voxels enclosed by the node atlas are extracted and fed into an autoencoder for dimension reduction (Fig~\ref{fig:brainnetworks}B). All node voxels are sampled and zero-padded to 3000 as the input of the autoencoder, which consists of 4 layers with dimensions 1024, 512, 128, 32, respectively, and the output dimension is 32. 

The IIT white matter atlas \cite{qi2021regionconnect} is a tractography atlas that indicates the path of 2227 the white matter tracts/edges connecting 68 regions of the Desikan atlas (Fig~\ref{fig:brainnetworks}C). Similar to nodes voxel extraction, the tract pathways of the IIT atlas are transformed back to the native space. Voxels of T1 and corresponding FA enclosed by IIT atlas are extracted and sampled as vectors (dimension 3000). Two multilayer perceptron (MLP, dimension: 3000, 1024, 512, 128, 32) respectively encode the voxel vectors of T1 and FA to features with a dimension of 32, and the project heads project the features to a common latent space (dimension = 128) where cross-modal contrastive representative learning are performed to align the features of T1 and FA (Fig~\ref{fig:brainnetworks}D). As such, the most tract-related features from T1 can be extracted under the guidance of the FA map. The loss is defined as:

\begin{equation}
    L = -log \frac{exp(cos(Z_{T1}(i),Z_{FA}(i))/\tau)}{\sum_{j=1}^{N} exp(cos(Z_{T1}(i),Z_{FA}(j))/\tau)}
\end{equation}

where $Z_{T1}$ and $Z_{FA}$ are latent features of T1 and FA features of the tract $i$ after the project head, respectively; $cos()$ is the cosine similarity; $\tau$ is the temperature parameter (set to 0.01) ;$N$ is the size of the minibatch, $j$ is the index of other tracts in the minibatch, $j\neq i$.

Finally, the node and edge features extracted from the trained models are reformatted into a graph for the downstream tasks (Fig~\ref{fig:brainnetworks}E).

\subsection{Learning framework}

\begin{figure}[hbt!]
\centering
\includegraphics[width=\textwidth]{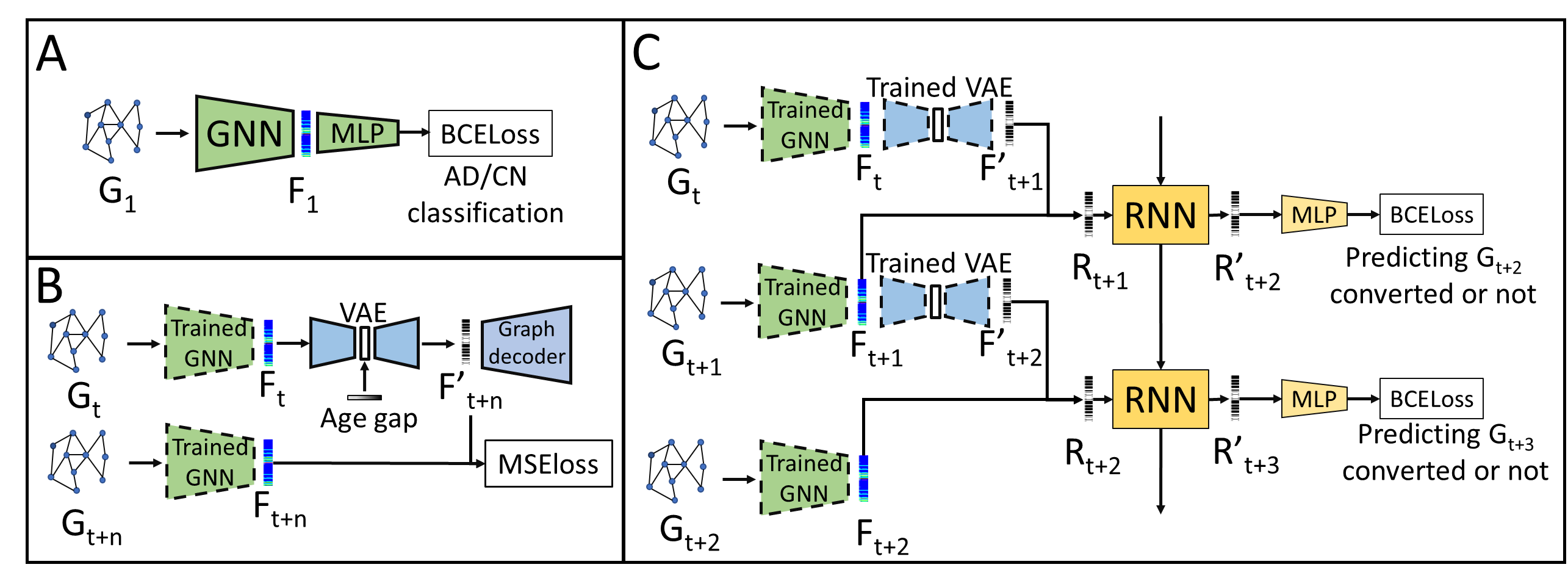}
\caption{Learning framework. \textbf{A}. A GNN is pretrained to extract brain network features $F$ by performing the AD/CN classification task. \textbf{B}.A VAE is pretrained with a longitudinal stable CN cohort to predict the changed brain network features due to the ageing effect by incorporating the age gap into the bottle neck of the VAE. A graph decoder decodes the predicted brain network for interpretation. \textbf{C}. An RNN is trained with the MCI cohort to perform the prediction of the MCI-to-AD conversion at $t+2$. based on the brain networks of $t$ and $t+1$} 
\label{fig:training}
\end{figure}

The learning framework for predicting MCI-to-AD conversion consists of three models that are trained separately. 

A GNN consisting of three GATConv~\cite{velivckovic2017graph} layers and one global pooling layer is pre-trained as graph encoder to extract dementia-related features $F_1$ from the brain networks $G_1$ by performing AD/CN classification on the independent baseline cohort with BCELoss (binary cross-entropy loss) (Fig~\ref{fig:training}A). The dimension of the output graph feature is 256. 

A VAE with a 3-layer (dimension: 256,64,16) MLP encoder and a 3-layer (dimension: 256,64,32) MLP decoder is trained to model the healthy ageing trajectories of the longitudinal stable CN cohort (Fig~\ref{fig:training}B). The task of the VAE is to predict the future features of brain networks $F_{t+n}$ by inputting a starting feature $F_{t}$ and age gap $n$. Specifically, the encoder projects the $F_{t}$ to a latent space which the age gap $n$ is fed into as a one-hot vector (dimension: 16). Then the decoder predicts the $F_{t+n}$ after the age-gap $n$ by training with the mean squared error loss (MSELoss) between predicted features $F'_{t+n}$ and the actual features $F_{t+n}$.

\begin{algorithm}
\caption{Training a RNN to predict conversion from MCI to AD}
\label{alg1}
\KwInput{Brain networks at $t$ and $t+1$: $\mathbf{G_t}$, $\mathbf{G_{t+1}}$} 
\For{$t = 1,2, \cdots$ }
{
Apply pretrained GNN: $F_t=\mathbf{GNN}(G_t), F_{t+1}=\mathbf{GNN}(G_{t+1})$ \\
Predict features at $t+1$ from $t$: $F'_{t+1}=\mathbf{VAE}(F_t)$ \\ 
Compute residuals: $R_{i+1} = F_{i+1} - F'_{i+1}$ \\
Predict $t+2$ residual with RNN:  $R'_{i+2} = \mathbf{RNN}(R_{i+1})$ \\
Predict MCI-to-AD conversion at $t+2$:  $P_{conv} = \mathbf{MLP}(R'_{i+2})$
}
\end{algorithm}

An RNN with long short-term memory (LSTM) kernel is trained to predict the MCI-to-AD conversion based on the longitudinal cohort containing both stable and converted MCI cohorts (Fig~\ref{fig:training}C). Briefly, the RNN predicts whether the conversion will happen in $t+2$ based on the brain networks of $G_t$ and $G_{t+1}$. The recurrent training details are explained in Algorithm~\ref{alg1}.

For interpretation purposes, a graph decoder with two MLP is trained with VAE: the first MLP (dimension: 256, 512, 1024, (68*32)) decodes node embedding from the brain network feature $F'_{t+n}$, and the second MLP (dimension: (32*2):64:128:64:32) decodes features of connecting edges between two nodes from the concatenated node embedding. 
%Our graph decoder is different from the conventional graph decoder that needs to predict the existence of an edge, as it only needs to perform regression on features of known edges. 
The residual between the predicted and actual graph features of brain networks represents the patient's deviation from the healthy ageing trajectories during the MCI-to-AD conversion. By comparing the predicted and actual brain networks that converted to AD, we could identify the abnormal changes that cannot be explained by the healthy ageing effect.

\subsection{Benchmarks}
To evaluate the performance of brain networks generated using our approach, we constructed traditional diffusion MRI and T1 based structural brain networks, respectively for comparison. 
For dMRI based brain networks (white matter connectivity), we performed whole-brain tractography on independent baseline CN cohort using the Anatomically-Constrained Tractography of MRtrix \cite{tournier2019mrtrix3}. The mean FA value of the tract fiber and the fiber counts is calculated as the edge weight among the nodes on the Desikan node atlas. 
For T1 based edge matrices (grey matter association), we measured the grey matter volumes constrained by the node atlas using FreeSurfer \cite{fischl2012freesurfer}. We calculated a covariance matrix to characterize the connectivity between brain regions.
All benchmark networks and our proposed networks were utilized in the classification of AD/CN of independent baseline cohort for evaluation using the graph encoder of the learning framework.

For the task of predicting MCI-to-AD conversion, we included two benchmarks representing static modal and dynamic model, respectively.
The static prediction benchmark is a residual neural network (ResNet) proposed in \cite{abrol2020deep}. Note that the original study predicted the conversion in 36 months, while we are predicting the conversion in 18 months. 
The interpretable temporal graph neural network (referred to as ITGNN) proposed in \cite{kim2021interpretable} was selected as the dynamic model benchmark, which consists of a GNN and an RNN. Briefly, the GNN encodes graph features of brain networks, and a LSTM learns from the graph feature to predict patients' outcomes (AD/CN/MCI). For benchmark purposes, we only included the longitudinal stable and converting MCI patients for training/testing and predict the patient status in 18 months. In addition, we input both the grey matter covariance networks and our proposed brain networks to produce two benchmark results.

All above models are implemented using Pytorch 1.10.0~\cite{paszke2019pytorch}. Five-fold cross validation, Adam optimiser, 0.0005 learning rate and 1000 training epochs and early convergence stopping are applied to all models.

\section{Results}

Results in table~\ref{tab:networks} show that our proposed method of generating brain networks achieved the highest performance for classifying AD/CN. Results in table~\ref{tab:conversion} show that our proposed method achieved highest prediction accuracy for the  MCI-to-AD conversion in 18 months .

\begin{table}[]
\centering
\caption{Performance of brain networks for AD/CN classification}
\label{tab:networks}
\begin{tabular}{cccc}
\hline
Models & Accuracy & Sensitivity & Specificity \\ \hline
Fibre Counts networks & 0.818 & 0.841 & 0.792 \\
FA networks & 0.828 & 0.850 & 0.802 \\
Cortical volume networks & 0.761 & 0.770 & 0.750 \\
Proposed networks (edge) & 0.842 & 0.858 & 0.822 \\
Proposed networks (edge + node) & \textbf{0.861} & \textbf{0.885} & \textbf{0.833} \\ \hline
\end{tabular}
\end{table}

\begin{table}[]
\centering
\caption{Performance comparison with benchmarks}
\label{tab:conversion}
\begin{tabular}{cccc}
\hline
Models & Accuracy & Sensitivity & Specificity \\ \hline
ResNet + baseline MRI & 0.802 & 0.813 & 0.794 \\
ITGNN+ cortical networks & 0.641 & 0.670 & 0.619 \\
ITGNN + proposed networks & 0.659 & 0.681 & 0.643 \\
Proposed + cortical networks & 0.779 & 0.791 & 0.770 \\
Proposed + proposed   networks & \textbf{0.839} & \textbf{0.868} & \textbf{0.818} \\ \hline
\end{tabular}
\end{table}

\begin{figure}[hbt!]
\centering
\includegraphics[width=0.45\textwidth]{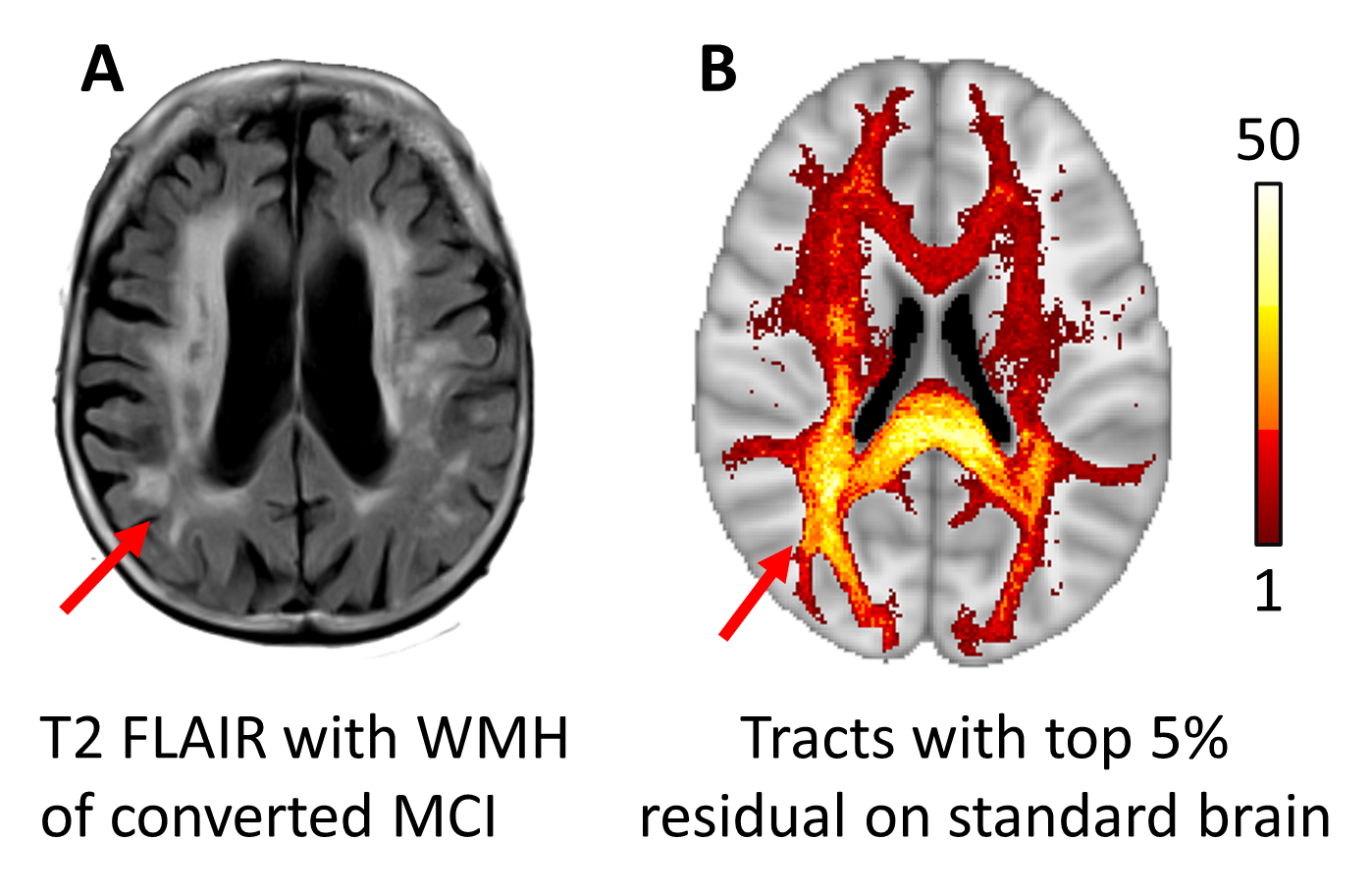}
\caption{Example of interpretation. \textbf{A.} A case example of a MCI patient who converted to AD. WMH(white matter hyper-intensity) is marked with red arrow. \textbf{B.} Distribution of tracts that are corresponding to the top $5\%$ residuals. Colorbar indicates number of edges crossing the voxel.}
\label{fig:interpretation}
\end{figure}

The interpretation approach produces a residual brain network with high dimensional features. For visualization purposes, we average the residuals of edge features, retain the edges with top 5$\%$ highest residuals, and present one case example in Fig~\ref{fig:interpretation}. The interpretation results suggest that the proposed methodology is capable of capturing the abnormalities in brain networks from MCI-to-AD conversion, particularly the white matter hyperintensities related to cognitive decline.rticularly the white matter hyperintensities related with cognitive decline.

\section{Discussion and conclusion}

This study proposes an approach to construct structural brain networks with imaging representation as the node and edge features and an approach to predict MCI-to-AD conversion based on the neuroscience knowledge that AD patients tend to deviate from healthy ageing trajectories. The proposed method outperforms benchmark methods. In addition, interpretation suggests that the proposed method is sensitive to abnormal structural changes in the brain. Future possible improvements include integrating separate training stages and introducing a quantitative model interpretation.  Overall, the proposed method shows promise to aid prognosis and risk assessment.

% ---- Bibliography ----
%
% BibTeX users should specify bibliography style 'splncs04'.
% References will then be sorted and formatted in the correct style.
%
\bibliographystyle{splncs04}
\bibliography{refs.bib}

\begin{thebibliography}{10}
\providecommand{\url}[1]{\texttt{#1}}
\providecommand{\urlprefix}{URL }
\providecommand{\doi}[1]{https://doi.org/#1}

\bibitem{abrol2020deep}
Abrol, A., Bhattarai, M., Fedorov, A., Du, Y., Plis, S., Calhoun, V.,
  Initiative, A.D.N., et~al.: Deep residual learning for neuroimaging: an
  application to predict progression to alzheimer’s disease. Journal of
  neuroscience methods  \textbf{339},  108701 (2020)

\bibitem{ajilore2014association}
Ajilore, O., Lamar, M., Kumar, A.: Association of brain network efficiency with
  aging, depression, and cognition. The American Journal of Geriatric
  Psychiatry  \textbf{22}(2),  102--110 (2014)

\bibitem{alexander2013convergence}
Alexander-Bloch, A., Raznahan, A., Bullmore, E., Giedd, J.: The convergence of
  maturational change and structural covariance in human cortical networks.
  Journal of Neuroscience  \textbf{33}(7),  2889--2899 (2013)

\bibitem{au2015back}
Au, R., Piers, R.J., Lancashire, L.: Back to the future: Alzheimer's disease
  heterogeneity revisited. Alzheimer's \& Dementia: Diagnosis, Assessment \&
  Disease Monitoring  \textbf{1}(3), ~368 (2015)

\bibitem{avants2009advanced}
Avants, B.B., Tustison, N., Song, G.: Advanced normalization tools (ants).
  Insight j  \textbf{2}(365),  1--35 (2009)

\bibitem{barile2021data}
Barile, B., Marzullo, A., Stamile, C., Durand-Dubief, F., Sappey-Marinier, D.:
  Data augmentation using generative adversarial neural networks on brain
  structural connectivity in multiple sclerosis. Computer methods and programs
  in biomedicine  \textbf{206},  106113 (2021)

\bibitem{cole2018brain}
Cole, J.H., Ritchie, S.J., Bastin, M.E., Hern{\'a}ndez, V., Mu{\~n}oz~Maniega,
  S., Royle, N., Corley, J., Pattie, A., Harris, S.E., Zhang, Q., et~al.: Brain
  age predicts mortality. Molecular psychiatry  \textbf{23}(5),  1385--1392
  (2018)

\bibitem{cox2019structural}
Cox, S., Ritchie, S., Fawns-Ritchie, C., Tucker-Drob, E., Deary, I.: Structural
  brain imaging correlates of general intelligence in uk biobank. Intelligence
  \textbf{76},  101376 (2019)

\bibitem{elliott2020mri}
Elliott, M.L.: Mri-based biomarkers of accelerated aging and dementia risk in
  midlife: How close are we? Ageing research reviews p. 101075 (2020)

\bibitem{fischl2012freesurfer}
Fischl, B.: Freesurfer. Neuroimage  \textbf{62}(2),  774--781 (2012)

\bibitem{gao2020ad}
Gao, F., Yoon, H., Xu, Y., Goradia, D., Luo, J., Wu, T., Su, Y., Initiative,
  A.D.N., et~al.: Ad-net: Age-adjust neural network for improved mci to ad
  conversion prediction. NeuroImage: Clinical  \textbf{27},  102290 (2020)

\bibitem{griffa2013structural}
Griffa, A., Baumann, P.S., Thiran, J.P., Hagmann, P.: Structural connectomics
  in brain diseases. Neuroimage  \textbf{80},  515--526 (2013)

\bibitem{gu2019generating}
Gu, X., Knutsson, H., Nilsson, M., Eklund, A.: Generating diffusion mri scalar
  maps from t1 weighted images using generative adversarial networks. In:
  Scandinavian Conference on Image Analysis. pp. 489--498. Springer (2019)

\bibitem{jenkinson2012fsl}
Jenkinson, M., Beckmann, C.F., Behrens, T.E., Woolrich, M.W., Smith, S.M.: Fsl.
  Neuroimage  \textbf{62}(2),  782--790 (2012)

\bibitem{kim2021interpretable}
Kim, M., Kim, J., Qu, J., Huang, H., Long, Q., Sohn, K.A., Kim, D., Shen, L.:
  Interpretable temporal graph neural network for prognostic prediction of
  alzheimer’s disease using longitudinal neuroimaging data. In: 2021 IEEE
  International Conference on Bioinformatics and Biomedicine (BIBM). pp.
  1381--1384. IEEE (2021)

\bibitem{li2021brainnetgan}
Li, C., Wei, Y., Chen, X., Sch{\"o}nlieb, C.B.: Brainnetgan: Data augmentation
  of brain connectivity using generative adversarial network for dementia
  classification. In: Deep Generative Models, and Data Augmentation, Labelling,
  and Imperfections, pp. 103--111. Springer (2021)

\bibitem{mattson2004pathways}
Mattson, M.P.: Pathways towards and away from alzheimer's disease. Nature
  \textbf{430}(7000),  631--639 (2004)

\bibitem{paszke2019pytorch}
Paszke, A., Gross, S., Massa, F., Lerer, A., Bradbury, J., Chanan, G., Killeen,
  T., Lin, Z., Gimelshein, N., Antiga, L., et~al.: Pytorch: An imperative
  style, high-performance deep learning library. Advances in neural information
  processing systems  \textbf{32} (2019)

\bibitem{qi2021regionconnect}
Qi, X., Arfanakis, K.: Regionconnect: Rapidly extracting standardized brain
  connectivity information in voxel-wise neuroimaging studies. Neuroimage
  \textbf{225},  117462 (2021)

\bibitem{sarasua2021transformesh}
Sarasua, I., P{\"o}lsterl, S., Wachinger, C., Neuroimaging, A.D., et~al.:
  Transformesh: A transformer network for longitudinal modeling of anatomical
  meshes. In: International Workshop on Machine Learning in Medical Imaging.
  pp. 209--218. Springer (2021)

\bibitem{schirmer2021neuropsychiatric}
Schirmer, M.D., Venkataraman, A., Rekik, I., Kim, M., Mostofsky, S.H., Nebel,
  M.B., Rosch, K., Seymour, K., Crocetti, D., Irzan, H., et~al.:
  Neuropsychiatric disease classification using functional connectomics-results
  of the connectomics in neuroimaging transfer learning challenge. Medical
  image analysis  \textbf{70},  101972 (2021)

\bibitem{song2020physics}
Song, T.A., Chowdhury, S.R., Yang, F., Jacobs, H.I., Sepulcre, J., Wedeen,
  V.J., Johnson, K.A., Dutta, J.: A physics-informed geometric learning model
  for pathological tau spread in alzheimer’s disease. In: International
  Conference on Medical Image Computing and Computer-Assisted Intervention. pp.
  418--427. Springer (2020)

\bibitem{tournier2019mrtrix3}
Tournier, J.D., Smith, R., Raffelt, D., Tabbara, R., Dhollander, T., Pietsch,
  M., Christiaens, D., Jeurissen, B., Yeh, C.H., Connelly, A.: Mrtrix3: A fast,
  flexible and open software framework for medical image processing and
  visualisation. Neuroimage  \textbf{202},  116137 (2019)

\bibitem{velivckovic2017graph}
Veli{\v{c}}kovi{\'c}, P., Cucurull, G., Casanova, A., Romero, A., Lio, P.,
  Bengio, Y.: Graph attention networks. arXiv preprint arXiv:1710.10903  (2017)

\bibitem{wei2021structural}
Wei, Y., Li, C., Cui, Z., Mayrand, R.C., Zou, J., Wong, A.L., Sinha, R., Matys,
  T., Sch{\"o}nlieb, C.B., Price, S.J.: Structural connectome quantifies tumor
  invasion and predicts survival in glioblastoma patients. bioRxiv  (2021)

\bibitem{wei2021quantifying}
Wei, Y., Li, C., Price, S.J.: Quantifying structural connectivity in brain
  tumor patients. In: International Conference on Medical Image Computing and
  Computer-Assisted Intervention. pp. 519--529. Springer (2021)

\end{thebibliography}

\end{document}